\documentclass{article}
\usepackage{amsfonts}
\usepackage{amsmath}
\usepackage{amssymb}
\usepackage{epsfig}
\usepackage{graphicx}

\begin{document}

\title{Long-range connections and mixed diffusion in fractional networks}
\author{\textit{R. Vilela Mendes}\thanks{%
rvilela.mendes@gmail.com, rvmendes@fc.ul.pt, http://label2.ist.utl.pt/vilela/%
} \\
CMAFCIO and IPFN, Universidade de Lisboa \and \textit{Tanya Ara\'{u}jo}%
\thanks{%
tanya@iseg.ulisboa.pt} \\
UECE, ISEG, Universidade de Lisboa}
\date{ }
\maketitle

\begin{abstract}
Networks with long-range connections obeying a distance-dependent power law
of sufficiently small exponent display superdiffusion, L\'{e}vy flights and
robustness properties very different from the scale-free networks. It has
been proposed that these networks, found both in society and biology, be
classified as a new structure, \textit{the fractional networks}. Particular
important examples are the social networks and the modular hierarchical
brain networks where both short- and long-range connections are present. The
anomalous superdiffusive and the mixed diffusion behavior of these networks
is studied here as well as its relation to the nature and density of the
long-range connections.
\end{abstract}

\section{Introduction}

The human brain contains up to 86 billion neurons connected by close to a
million kilometers of axons and dendrites. Most of these connections ($\sim $%
80\%) are short range on the order of a few hundred microns, the rest ($\sim 
$20\%) being long-range myelinated fibers on the order of several
centimeters. The insulating myelin sheath increases conduction velocity of
the action potentials but at the cost of taking up more volume in the brain
as well as rendering axons unable to synapse onto nearby neurons. That
evolution has found profitable to accept this additional hardware cost
highlights the importance of long-range connections.

From a network point of view the brain has a modular and hierarchical
structure \cite{Park} \cite{Wig}. Each module is associated to a specialized
function mediated by short-range connections whereas global integration, for
higher cognition functions, relies on the long-range connections between
modules.

The existence and importance of long-range connections in the brain has been
much studied in recent years \cite{Knosche} \cite{Betzel} \cite{Padula} \cite%
{Markov} \cite{Modha} \cite{Fluo}, with diminished long-range functional
connectivity being associated to cognitive disorders \cite{Barttfeld}. Of
course, by itself, existence of long-range connections between the
specialized nodes does not guarantee global integration of the cognitive
functions. It is also necessary that the flow of information be sufficiently
fast for the stimulus integration to be performed in a timely manner. This
seems of particular relevance for the forward and backwards loops in the
predictive coding mode \cite{Clark1} \cite{Friston2} \cite{Friston3} \cite%
{Clark2} \cite{Spratling} \cite{Hogendoorn} of brain operation. One may
therefore ask what type of communication short- and long-range connections
establish and whether it depend or not on the structure and density of the
long-range connections.

The network modules in the brain are in fact repertoires of many neurons
and, when dealing with the interactions of these intrinsic connectivity
networks (ICN's), a continuous diffusion approximation might be a good
modelling hypothesis. In another paper \cite{Vilela-FNL} the nature of the
diffusion processes associated to short and long-range connections have been
analyzed. In particular it was concluded that whereas for short-range
connections information propagates as a normal diffusion, for long-range
connections of a certain type, one has anomalous diffusion, sub- or
super-diffusion depending on the power law distance dependence of the
connections.

Networks with long-range connections leading to superdiffusion display
properties so very different from scale-free and hub dominated networks that
in \cite{Vilela-FNL} it was proposed to characterize them as a new class of
networks, \textbf{the fractional networks}. Notice that long-range
connections are also important in social networks \cite{Hogan} \cite%
{Romantic} \cite{Carvalho} \cite{Gustafson}.

In a network the Laplacian matrix is%
\begin{equation}
L=G-A  \label{1.1}
\end{equation}%
$G$ being the degree matrix ($G_{ij}=\delta _{ij}\times $ number of
connections of node $i$) and $A$ the adjacency matrix ($A_{ij}=1$ if $i$ and 
$j$ are connected, $A_{ij}=0$ otherwise). Let $\psi \left( i\right) $ for
each node $i$ be the intensity of some function $\psi $ across the network.
For a node $i$ connected along some coordinate to two other nearest neighbor
nodes $i+1$ and $i-1$ the action of the Laplacian matrix on a vector leads
to $-\psi \left( i-1\right) +2\psi \left( i\right) -\psi \left( i-1\right) $%
, which is a discrete version of $-d^{2}$ (minus the second derivative). It
is reasonable to think that $\psi $ diffuses from $i$ to $j$ proportional to 
$\psi \left( i\right) -\psi \left( j\right) $ if $i$ and $j$ are connected.
Then,%
\begin{equation}
\frac{d\psi \left( i\right) }{dt}=-k\sum_{j}A_{ij}\left( \psi \left(
i\right) -\psi \left( j\right) \right) =-k\left( \psi \left( i\right)
\sum_{j}A_{ij}-\sum_{j}A_{ij}\psi \left( j\right) \right)  \label{1.2}
\end{equation}%
which in matrix form is%
\begin{equation}
\frac{d\psi }{dt}+kL\psi =0  \label{1.3}
\end{equation}%
a heat-like equation. Therefore the Laplacian matrix controls the diffusion
of quantities in the network and in the continuous approximation and for
short-range connections the propagation of signals in the network may be
represented by a normal diffusion equation.%
\begin{equation}
\frac{d\psi }{dt}=k\Delta \psi  \label{1.4}
\end{equation}%
$\Delta $ being the Laplacian in the dimension of the space where the
network is embedded.

However, for long-range connections the situation is different and from the
symmetrized Gr\"{u}nwald-Letnikov representation of the fractional
derivative it was found \cite{Vilela-FNL} (see also the Appendix) that for
networks where the probability of establishment of a link at distance $d$ is
proportional to a power of the distance%
\begin{equation}
P_{ij}=cd_{ij}^{-\gamma }  \label{1.5}
\end{equation}%
diffusion would be fractional diffusion of exponent $\beta =\gamma -1$. $%
\beta =2$ being normal diffusion and all $\beta <2$ corresponding to
superdiffusions.%
\begin{equation}
\frac{d\psi }{dt}=-k\left( -\Delta \right) ^{\frac{\beta }{2}}\psi
\label{1.6}
\end{equation}

Anomalous diffusion and other phenomena \cite{Vilela-FNL} emerge naturally
as a structural property in long-range connection networks with distance
dependence as in (\ref{1.5}). Here, in Section 2, the case of networks
characterized by a modular hierarchical structure with both short and long
range connections will be studied. This is the structure that occurs in
brain networks and also in some social networks. Whereas in the networks
studied in \cite{Vilela-FNL} the uniform scaling law of the connections
leads to pure anomalous diffusion, here one faces a mixture of both normal
and anomalous diffusion. This is the central phenomena that is studied in
this paper with emphasis on the nature of the time scales of propagation of
information. This is discussed in the framework of the continuous
approximation to the network leading to a fractional differential equation.
The continuous approximation is a reasonable approximation for very large
networks. However it is also found that qualitatively similar results are
obtained even for small discrete networks. This is illustrated in Section 3,
by numerical simulation in a relatively small network (400 nodes).

\section{Mixed diffusion}

In the mixed case the diffusion equation will be%
\begin{equation}
\frac{d\psi \left( x,t\right) }{dt}=\left( a\Delta -b\left( -\Delta \right)
^{\frac{\beta }{2}}\right) \psi \left( x,t\right)  \label{2.1}
\end{equation}%
with $x\in \mathbb{R}^{n}$, $n$ being the dimension of the embedding
Euclidean space. For the Fourier transform%
\begin{equation}
\widetilde{\psi }\left( k,t\right) =\int d^{n}x\psi \left( x,t\right)
e^{-ik\cdot x}  \label{2.2}
\end{equation}%
one has the equation%
\begin{equation}
\frac{d\widetilde{\psi }\left( k,t\right) }{dt}=\left( -a\left\vert
k\right\vert ^{2}-b\left\vert k\right\vert ^{\beta }\right) \widetilde{\psi }%
\left( k,t\right)  \label{2.3}
\end{equation}%
with solution%
\begin{equation}
\widetilde{\psi }\left( k,t\right) =\widetilde{\psi }\left( k,0\right)
e^{-t\left( a\left\vert k\right\vert ^{2}+b\left\vert k\right\vert ^{\beta
}\right) }  \label{2.4}
\end{equation}%
$\widetilde{\psi }\left( k,0\right) =1$ corresponds to $\psi \left(
x,0\right) =\delta ^{\left( n\right) }\left( x\right) $, that is, an initial
localized disturbance at the origin. This is the situation of interest to
study the propagation of information in the network. Computing the inverse
Fourier transform one has%
\begin{equation}
\psi \left( x,t\right) =\frac{2A_{n}}{\left( 2\pi \right) ^{n-1}}%
\int_{0}^{\infty }d\left\vert k\right\vert \left\vert k\right\vert
^{n-1}e^{-t\left( a\left\vert k\right\vert ^{2}+b\left\vert k\right\vert
^{\beta }\right) }\frac{\sin \left( \left\vert k\right\vert \left\vert
x\right\vert \right) }{\left\vert k\right\vert \left\vert x\right\vert }
\label{2.5}
\end{equation}%
with%
\begin{equation}
A_{n}=\left\{ 
\begin{array}{l}
\frac{\pi ^{m-1}2^{m-2}}{\left( 2m-2\right) !!}\;\;n=2m \\ 
\frac{\pi ^{m-1}2^{m-1}}{\left( 2m-1\right) !!}\;\;n=2m+1%
\end{array}%
\right.  \label{2.6}
\end{equation}%
As in the purely fractional multidimensional solution \cite{Hanyga} one
notices the strong dependence on the dimension $n$.

\begin{figure}[htb]
    \centering
    \includegraphics[width=0.5\textwidth]{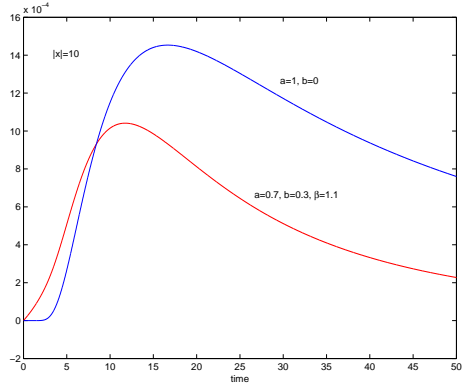}
    \caption{Comparison of the propagation time of a delta signal at $\left( x=0,t=0\right) $ to a distance $x=10$ for normal and mixed diffusion ($\protect\beta =1.1,a=0.7,b=0.3$)}
    \label{d_10}
\end{figure}

\begin{figure}[htb]
    \centering
    \includegraphics[width=0.5\textwidth]{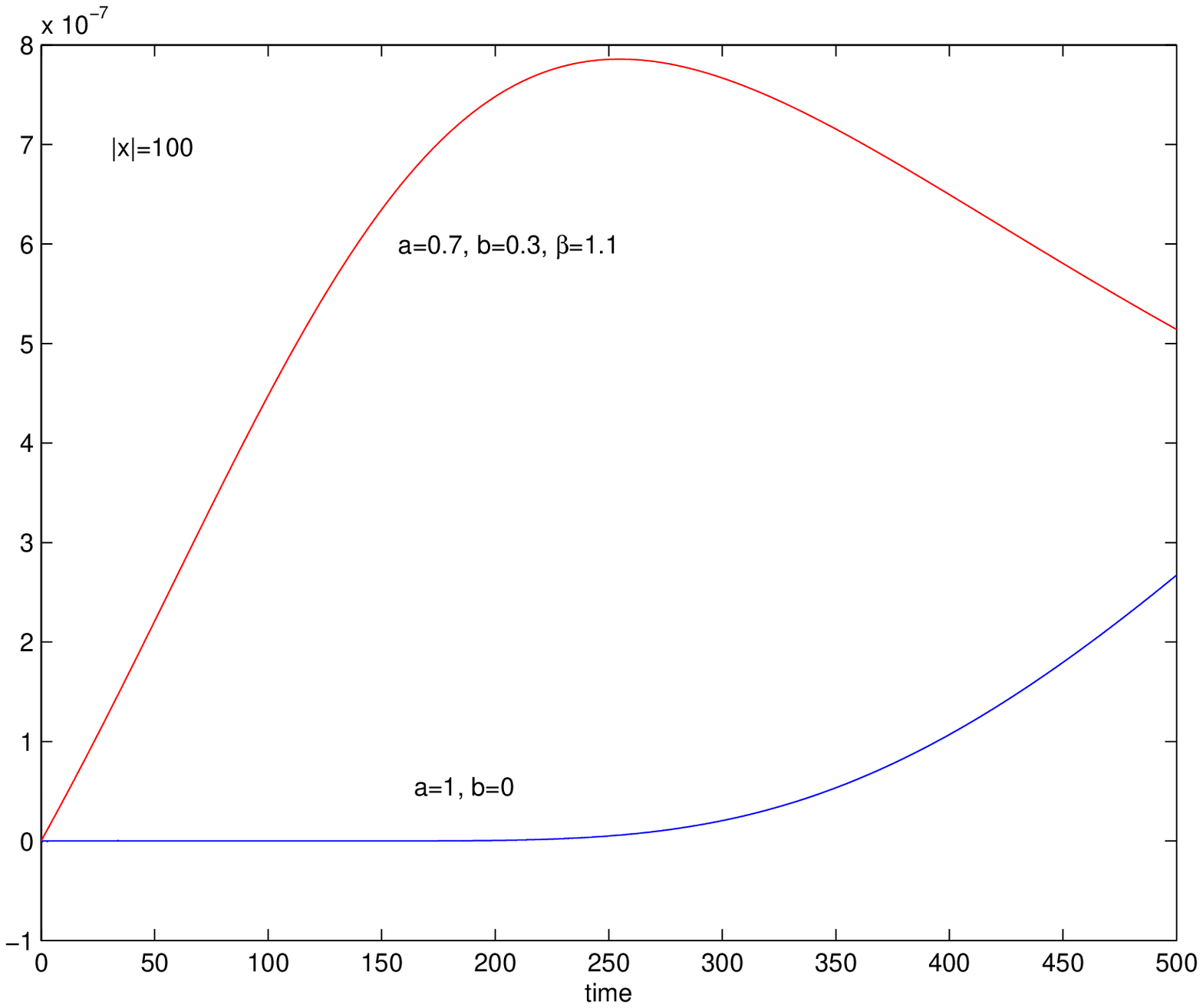}
    \caption{Comparison of the propagation
time of a delta signal at $\left( x=0,t=0\right) $ to a distance $x=100$ for
normal and mixed diffusion ($\protect\beta =1.1,a=0.7,b=0.3$)}
    \label{d_100}
\end{figure}

Numerical evaluation of (\ref{2.5}) shows the remarkable difference in the
speed of propagation of information between normal and mixed diffusion. For $%
n=3$, Figures \ref{d_10} and \ref{d_100} compare the propagation of a delta
signal at $\left( x=0,t=0\right) $ to distances $x=10$ and $100$ for normal
and mixed diffusion. One sees that whereas for normal diffusion it takes a
long time for the signal to be detected at a distance, for mixed diffusion
the behavior is qualitatively very different.

\begin{figure}[htb]
    \centering
    \includegraphics[width=0.5\textwidth]{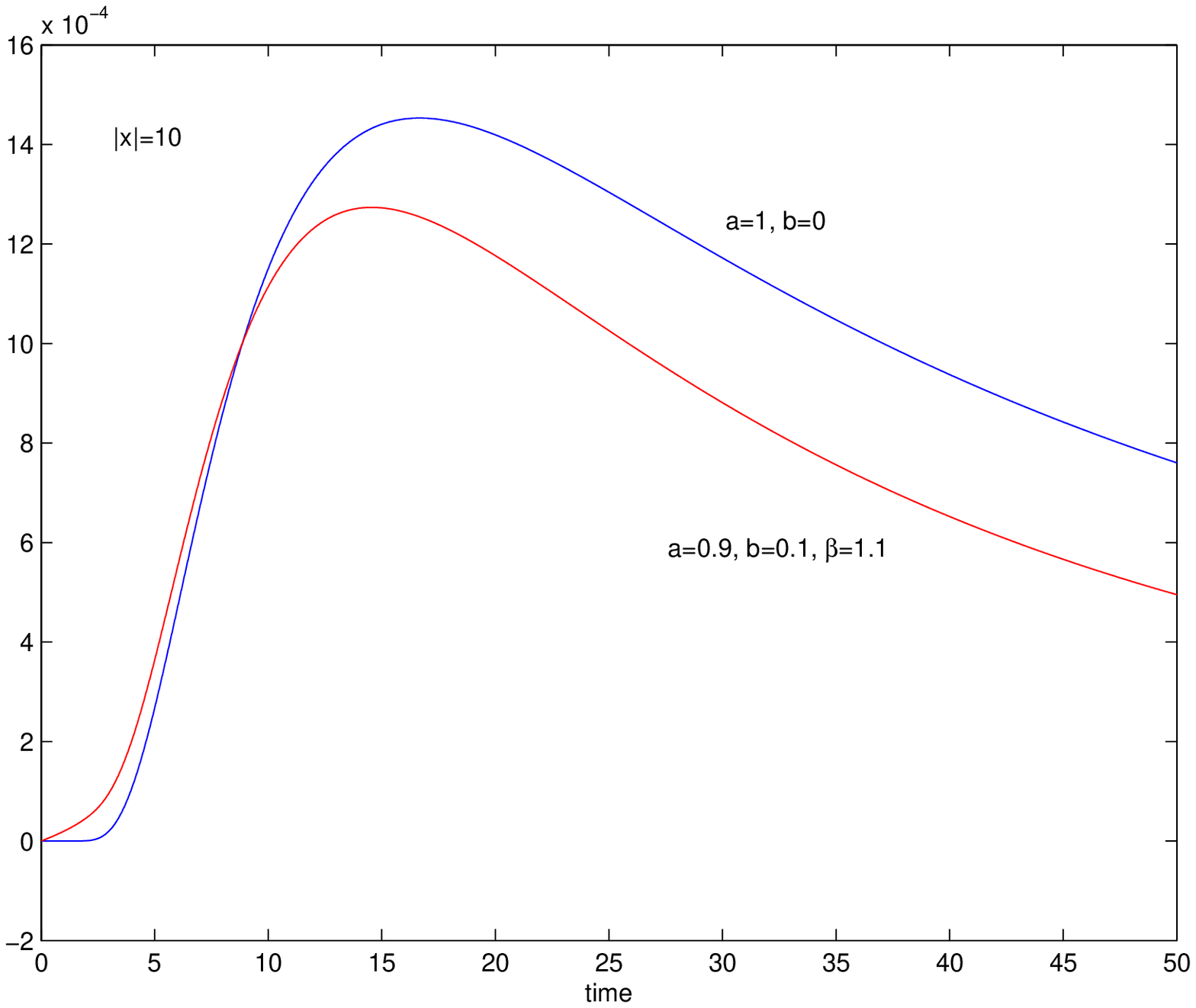}
    \caption{Comparison of the propagation
time of a delta signal at $\left( x=0,t=0\right) $ to a distance $x=10$ for
normal and mixed diffusion ($\protect\beta =1.1,a=0.9,b=0.1$)}
    \label{d_10_2}
\end{figure}

\begin{figure}[htb]
    \centering
    \includegraphics[width=0.5\textwidth]{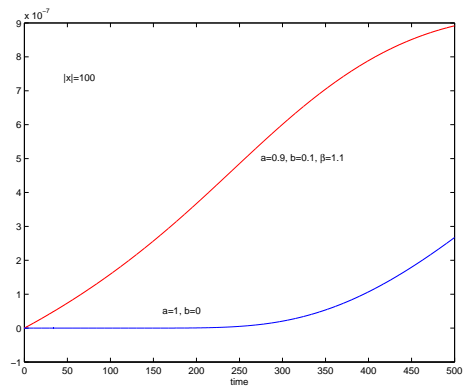}
    \caption{Comparison of the propagation
time of a delta signal at $\left( x=0,t=0\right) $ to a distance $x=100$ for
normal and mixed diffusion ($\protect\beta =1.1,a=0.9,b=0.1$)}
    \label{d_100_2}
\end{figure}

Figures \ref{d_10_2} and \ref{d_100_2} show that this effect is obtained
even with a very small amount of fractional diffusion.

Of course the effect exist only if $\beta <2$. For $\beta \geq 2$ the behavior
would be practically indistinguishable from normal diffusion. This puts into
evidence the fact that the mere existence of long-range connections does not
guarantee the existence of fractional superdiffusion. That is, a sufficient
density of long-range connections to be at least consistent with the one in (%
\ref{1.5}) is required. This is an important hint to be taken into account
on the relation of functional connectivity to brain cognitive disorders.
\clearpage

\section{Diffusion in a small fractional network: Numerical results}

So far we have discussed the diffusion behavior of fractional networks in
the framework of the continuous approximation to the network. Here, by
numerically simulating the propagation of a pulse of information in a
discrete network, we show that the results are consistent with those
obtained from the continuous approximation modeled by the fractional
differential equations.

\begin{figure}[htb]
    \centering
    \includegraphics[width=0.7\textwidth]{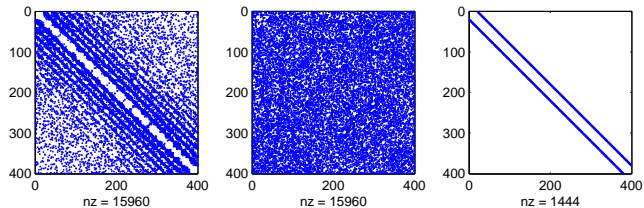}
    \caption{Connection patterns
for a fractional network with $\gamma =2$ and sparsity $0.1$, for a
random network with the same sparsity and a nearest-neighbor network}
    \label{Adjs}
\end{figure}

We consider 400 agents (nodes) placed in two-dimensional 20x20 grid and
establish connections among the nodes with a distance-dependent power-law
distribution%
\begin{equation*}
P_{ij}\symbol{126}d^{-\gamma }
\end{equation*}%
Namely, we pick a node at random and establish a connection to another node
at a distance $d$%
\begin{equation*}
d=\exp \left\{ \frac{\log \left( d_{\min }^{1-\gamma }-C\gamma y\right) }{%
1-\gamma }\right\}
\end{equation*}%
$y$ being a random number in the interval $\left[ 0,1\right] $ and $C$ a
constant%
\begin{equation*}
C=\frac{\left( d_{\min }^{1-\gamma }-d_{\max }^{1-\gamma }\right) }{\gamma }
\end{equation*}%
In Fig.\ref{Adjs} we show the pattern of connections for the fractional
network with $\gamma =2$ and a sparsity index of $0.1$. In the same figure
are also shown the patterns of connections for a random network with the
same sparsity and for a nearest-neighbor network with the maximal number of
connections.

\begin{figure}[htb]
    \centering
    \includegraphics[width=0.7\textwidth]{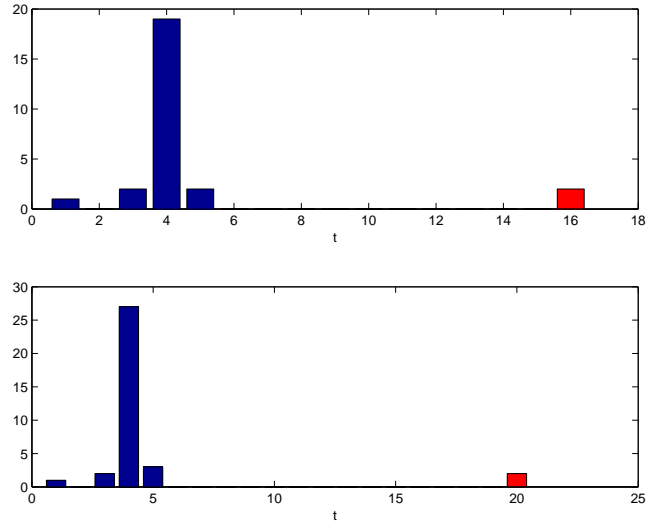}
    \caption{Propagation of a unit pulse
between two distant nodes for a fractional network $\left( \protect\gamma %
=2\right) $, compared with the same phenomenon in a nearest-neighbor network
(the single pulses at t=16 and t=20)}
    \label{signals}
\end{figure}

To study the diffusion in the fractional network, we consider, at time zero,
a unit pulse at one node and study how it propagates throughout the network.
At each time step the pulse is transmitted to the neighbors of each
activated node, with a no-backflow condition being imposed. That is, after a
node transmits the pulse to its neighbors it no longer transmits the same
pulse even if it receives it back through some cycle in the network. In the
Fig.\ref{signals} we show the results of two typical simulations. In each
case we have chosen, among the nodes that have a long-distance connection,
those that are further apart. In the same figures we compare with the
results of the same experiment for a nearest-neighbor network (the single
pulses at times 16 and 20). Not only is the signal transmitted much faster
in the fractional network, but also its coherent nature is preserved,
instead of being spread over a very large number of distinct times as it may
occur in a sparse random network. Very similar behavior is also obtained for
the propagation between nodes that are not directly connected at time one.

\clearpage

\section{Remarks and conclusions}

1. As has been experimentally confirmed, existence of long-range connections
between the brain ICN's is critical for integration and interpretation of
sensory stimuli and higher cognitive functions. One view of brain
integration and consciousness \cite{Zhou} \cite{Enzo} is based on a
percolation model. For percolation, that is, for the formation of a global
cluster, it suffices that connections exist between the local clusters.
However for the establishment of higher cognitive functions, and in
particular in the predictive coding mode, it is necessary that the
interaction between the ICN's be established at a sufficiently fast rate.
Therefore the mere existence of long-range connections is not sufficient, it
also necessary that they have, for example, a power-law dependence with $%
\gamma <3$.

2. The additional hardware cost of myelinated long-range connections in the
brain is compensated by the integration of information and higher cognitive
functions. Another puzzling additional energetic cost is that, when tested
with fMRI, the resting brain is in fact turbulent and restless \cite{Raichle}%
. There is a good reason for that, probably related to speed of reaction.
With the operating time scales of individual neurons and their low average
firing rate, pattern recognition by evolution towards an equilibrium fixed
point or minimizing an energy function would be much too slow for practical
living purposes. As has been conjectured, for example from the studies of
the olfactory bulb \cite{Freeman}, a much faster recognition is achieved by
replacing the low-level chaos that exists in the absence of an external
stimulus by, in the presence of a signal, a pattern of bursts with different
intensities in different regions. A network of Bernoulli units \cite{Dente}
is a model confirmation of this conjecture.

3. Finally, as already discussed in \cite{Vilela-FNL}, the robustness and
controllability properties of the fractional networks are so very different
from the scale-free networks that they deserve a detailed study. This is
relevant not only for brain functions but also concern the uses and misuses
of information flow in the social networks.

\section*{Appendix: Power-law long-range connections and fractional diffusion}

For completeness we include here a short derivation of the relation between
power-law long-range connections and fractional diffusion equations, already
discussed in Ref.\cite{Vilela-FNL}.

Let the probability of a link at distance $d$ be proportional to a power of
the distance%
\begin{equation*}
P_{ij}=cd_{ij}^{-\gamma }\mathnormal{\hspace{2cm}}\text{\textnormal{\ with }}%
\gamma \leq 3
\end{equation*}%
Consider now a block renormalized network $N^{\ast }$ where each set of $q$
nearby nodes in the original network $N$ are mapped to a node of the $%
N^{\ast }$ network. With the block renormalization, the power-law connection
probability leads to actual connection strengths in the renormalized
network. In the $N^{\ast }$ network the connections are%
\begin{equation*}
A_{ij}^{\ast }\simeq cqd_{ij}^{-\gamma }
\end{equation*}%
with the Laplacian $L^{\ast }$ and degree$\ G^{\ast }$ matrices of the $%
N^{\ast }$ network being 
\begin{equation*}
L^{\ast }\psi \left( i\right) =G_{ii}^{\ast }\psi \left( i\right)
-cq\sum_{j\neq i}d_{ij}^{-\gamma }\psi \left( j\right)
\end{equation*}%
Compare the distance dependence of the elements of the Laplacian matrix $%
L^{\ast }$ along one of the coordinate axis with a discrete one-dimensional
representation of a fractional derivative. The symmetrized Gr\"{u}%
nwald-Letnikov representation of the fractional derivative $\left(
a<x<b\right) $ (see \cite{Mainardi}) is%
\begin{eqnarray}
D^{\beta }\psi \left( x\right) &=&\frac{1}{2}\lim_{h\rightarrow 0}\frac{1}{h}%
\left\{ \sum_{n=0}^{\left[ \frac{x-a}{h}\right] }\left( -1\right) ^{n}\left( 
\begin{array}{c}
\beta \\ 
n%
\end{array}%
\right) \psi \left( x-nh\right) \right.  \notag \\
&&\left. +\sum_{n=0}^{\left[ \frac{b-x}{h}\right] }\left( -1\right)
^{n}\left( 
\begin{array}{c}
\beta \\ 
n%
\end{array}%
\right) \psi \left( x+nh\right) \right\}  \label{A1}
\end{eqnarray}%
with coefficients%
\begin{equation}
\left\vert \left( 
\begin{array}{c}
\beta \\ 
n%
\end{array}%
\right) \right\vert =\frac{\Gamma \left( \beta +1\right) \left\vert \sin
\left( \pi \beta \right) \right\vert }{\pi }\frac{\Gamma \left( n-\beta
\right) }{\Gamma \left( n+1\right) }\underset{n\text{ large}}{\backsim }%
\frac{\Gamma \left( \beta +1\right) \left\vert \sin \left( \pi \beta \right)
\right\vert }{\pi }n^{-\left( \beta +1\right) }  \label{A2}
\end{equation}%
and $sign\left( 
\begin{array}{c}
\beta \\ 
n%
\end{array}%
\right) =\left( -1\right) ^{n+1}$.

Comparing (\ref{A1}-\ref{A2}) with the expression for $L^{\ast }\psi \left(
i\right) $, the conclusion is that diffusion in the $N^{\ast }$ network is
fractional diffusion of exponent $\beta =\gamma -1$. $\beta =2$ would be
normal diffusion, all $\beta <2$ corresponding to superdiffusions.

\end{document}